\documentstyle[prb,aps,epsf]{revtex}

\newcommand{\ie}{{\em i.e.}}

\newcommand{\eq}[1]{Eq.~(\ref{#1})}
\newcommand{\peq}{P^{\rm eq}}

\begin{document}
\draft

\twocolumn[\hsize\textwidth\columnwidth\hsize\csname@twocolumnfalse%
\endcsname

\title{
Convergence of Monte Carlo Simulations to Equilibrium
}

\author{Onuttom Narayan\cite{newaddress} and A. P. Young\cite{byline}}
\address{Department of Physics, University of California, Santa Cruz,
CA 95064}

\date{\today}

\maketitle

\begin{abstract}
We give two direct, elementary proofs that a Monte Carlo simulation converges
to equilibrium provided that appropriate conditions are satisfied.
The first proof requires detailed balance while the second is quite
general.
\end{abstract}

%\vskip 0.3 truein
]

Monte Carlo simulations are widely used in statistical
physics.
If the algorithm satisfies the detailed balance condition,
it is easy to show that
the desired
%distribution~\cite{gen}
distribution
is a {\em stationary}\/ distribution, i.e. 
if the system is, by some means, put into
the desired
distribution, it will subsequently stay in this distribution.
It is harder, but nonetheless crucial, to also show that, starting from a
general distribution, the algorithm will {\em
converge}\/ to the desired distribution.
Although this can be proved without too much difficulty for a system with 
a finite number of states~\cite{K1}, this
proof is unfamiliar to most physicists. The argument in the
original paper of Metropolis et al.\cite{metropolis} makes convergence
plausible but is not a proof\cite{plausible}.
Most physics texts on Monte Carlo methods
either do not give a proof of
convergence\cite{nb-binder2} or refer the reader\cite{binder,sokal} to
rather abstract derivations in the mathematics literature\cite{doob}, or
rely on the Frobenius-Perron theorem\cite{sokal} which is unfamiliar to most
physicists.

In this paper we present two proofs of convergence which are
self contained
and use only elementary methods.  We feel that it is useful to present these
derivations here because (i) it is not widely known in the physics community
that it is not difficult
to prove convergence, at least for systems with a finite
number of states, and 
(ii) our proofs are (to the best of our knowledge) different
from and as simple as existing proofs in the mathematical
literature. Even the proof of Ref.\onlinecite{K1}, which is of comparable simplicity,
is hard to understand physically; this is discussed more fully near the end of 
the paper.

The  first proof relies on the Monte Carlo algorithm satisfying the condition
of {\it detailed balance.\/} Although this is true in essentially all Monte
Carlo simulations, it is not
%an essential requirement
strictly required
for the algorithm to
converge to the equilibrium distribution. 
In the second half of this paper, we
shall also present  a more general proof which relaxes this condition, and which
is very different from the proof assuming detailed balance. 

Throughout this paper, we shall assume that the system being considered has a
finite number of states. Most systems in physics can be approximated as such
by discretizing any continuous variables sufficiently finely; for instance, in
molecular simulations, it should be permissible to limit the phase space for
any particle to a sufficiently large region, and then discretize it in small
intervals.  A rigorous proof of convergence for systems with an infinite
number of states is much more complicated~\cite{Chung}.

In the simplest case, the desired distribution that the Monte Carlo method
seeks to simulate is the Boltzmann distribution at some temperature.
However, in dealing with glassy systems, it is sometimes more efficient to
simulate a different distribution, as in the
multicanonical ensemble\cite{bc},
the $1/k$ ensemble\cite{hs}, and parallel tempering\cite{hn}. The results
presented in this paper are valid whenever the desired distribution is a
stationary distribution of the algorithm, and detailed balance --- or,
more generally, ergodicity --- is satisfied. This includes the cases
mentioned above.

The essential ingredients of the Monte Carlo method in statistical physics are
the (non-negative) ``transition rates'',
$w_{l\rightarrow m}$\,, defined to be
the probability that, given the system is in
state $l$ at ``time'' $t$, then it will be in state $m (\ne l)$
at time $t+1$. We define time to be incremented by one every Monte Carlo move
(not sweep) and assume initially that all moves are equivalent, so the
$w_{l\rightarrow m}$ do not depend on time. An example would be
flipping a single spin chosen at random. The important case of sequential
updating will be discussed later.

The probability that the system is in state $l$ at time $t$ is defined to be
$P_l(t)$.
The evolution of these probabilities is governed by the ``master
equation'',
\begin{equation}
P_l(t+1) - P_l(t) =
\sum_{m \ne l}
\left[ P_m(t)\, w_{m\rightarrow l}\, - P_l(t)\, w_{l\rightarrow m} \,
\right].
\label{master}
\end{equation}
The first term on the right hand side describes transitions into state $l$ from
$m$ (which therefore increases $P_l$ and so has a plus sign) while the second
term describes transitions out of state $l$, which decreases $P(l)$. Note that
only terms with $m \ne l$ contribute. We can also define 
$w_{l\rightarrow l}$ to be the probability that the system stays in state $l$,
i.e. $w_{l\rightarrow l} = 1 - \sum_{m \ne l} w_{l\rightarrow m}$, or
equivalently,
\begin{equation}
\sum_{m } {w_{l\rightarrow m}}  =  1 .
\label{constraints}
\end{equation}
Eq.~(\ref{constraints}) implies that the master equation can be written
\begin{equation}
P_l(t+1) = \sum_{m}
P_m(t)\, w_{m\rightarrow l} ,
\label{master2}
\end{equation}
where the term $m = l$ is now included.

A necessary  condition for the method to work is that the desired
distribution, $\peq$,
is {\em stationary}, \ie\ if $P_l(t) = \peq_l$ for all $l$ then
$P_l(t+1)=\peq_l$. 
This means that the right hand side of Eq.~(\ref{master})
must vanish for $P=\peq$. 
The condition of {\it detailed balance\/} consists of the 
assumption that
{\em each} term on
the right hand side of \eq{master} separately vanishes for $P=\peq$, \ie\
\begin{equation}
\peq_l\, w_{l\rightarrow m}\, = \peq_m\, w_{m\rightarrow l}\,.
\label{db}
\end{equation}
In the first part of this paper, we shall assume that Eq.~(\ref{db}) is
satisfied.

We start with the following quantity, which is a measure of the deviation from
equilibrium,
\begin{equation}
G = \sum_l {1 \over \peq_l} \left(P_l - \peq_l \right)^2 
= \sum_l \left( {P_l^2 \over \peq_l}\right) - 1  
\label{quaddev}
\end{equation}
evaluated at time $t$, where
the last expression follows because
$P$ and $\peq$ are normalized.

At time $t+1$ we indicate (for compactness of notation)
the probabilities by $P^\prime_l$ and the corresponding value of $G$ by
$G^\prime$. We will show that $G$ monotonically decreases, \ie\
\begin{equation}
\Delta G \equiv G^\prime - G \leq 0 ,
\label{ineq}
\end{equation}
where the equality only holds if $G$ and $G^\prime$ both vanish, so the system
is in equilibrium. This shows that
the system will
eventually approach arbitrarily close to the equilibrium distribution.

Using Eqs.~(\ref{master2}) and (\ref{quaddev}), $\Delta G$ can be written as
\begin{equation}
\Delta G = \sum_{l, m, n} \left[ w_{m\rightarrow l} w_{n\rightarrow l} {
P_m P_n \over \peq_l} \right]
- \sum_l {P_l^2 \over \peq_l} .
\label{dg}
\end{equation}
In the first term on the right hand side of Eq.~(\ref{dg}) we use the detailed
balance condition, Eq.~(\ref{db}), to replace $w_{m\rightarrow
l}$ by $w_{l\rightarrow m} \peq_l/\peq_m$, and in the second term we can
use Eq.~(\ref{constraints}) to  insert
a factor of $\sum_{m } w_{l\rightarrow m}$ (and interchange the indices $l$
and $m$). This gives
\begin{equation}
\Delta G = \sum_{l, m, n}  \left[ w_{l\rightarrow m} w_{l\rightarrow n} 
\peq_l {P_m P_n \over \peq_m \peq_n} \right] 
-\sum_{l, m} w_{m\rightarrow l} {P_m^2 \over \peq_m} 
\end{equation}
Applying the detailed balance relation again and incorporating a factor of
$\sum_n w_{l\rightarrow n}$,
the last term in the above equation can be written as
\begin{equation}
-\sum_{l, m, n} w_{l\rightarrow m}
w_{l\rightarrow n} \peq_l  \left({P_m \over \peq_m}\right)^2 .
\end{equation}
Taking the half the sum of this and the same expression with $m$
replaced by $n$, we finally get
\begin{equation}
\Delta G =  
-{1\over 2}\sum_{l,m,n}  w_{l\rightarrow m} w_{l\rightarrow n}\, \peq_l 
\left( {P_m \over \peq_m} - {P_n \over \peq_n} \right)^2   ,
\label{dg_res}
\end{equation}
where terms with $m = l$ and $n=l$ are included. 

Eq.~(\ref{dg_res}) is the 
main result of this part of the paper.
It shows that $\Delta G$
is definitely negative unless, for every state $l$, all states 
which can be reached from $l$ in a single move --- equivalently, with detailed
balance, all states from which $l$ can be reached in a single
move --- have probabilities 
proportional to the equilibrium probabilities. 
The most natural scenario is that {\em all}\/ states satisfy this
with the same proportionality constant (which must be unity)
i.e. the
system is in equilibrium. However, $\Delta G$ also vanishes if 
%We have therefore
%proved the desired result, \eq{ineq}.
%% with the equality only holding if the system is in equilibrium.  
$P_m/\peq_m$ assumes different values for states which have no common
one-step descendants. Hence, to achieve full equilibrium,
the algorithm must also
be ergodic, \ie\ starting from a given state, after a sufficiently long time
there is non-zero probability for the system to be in any state.
The condition of ergodicity is sufficient to ensure that even if 
$\Delta G$ is accidentally zero at some time step, it must decrease
later, since any two states must have common descendants after several
time steps. 
If in addition $w_{l\rightarrow l}\neq 0$ for all $l,$ which is
usually
true, the one-step descendants of a set of states must include the set 
itself, so that it is not possible to break up all the states of the
system into subsets
with no common one-step descendants across two subsets. Thus if this 
condition is satisfied, $\Delta G$ cannot be zero (without the system
being in equilibrium) at {\em any\/} time step. 

We will distinguish between a process which is ergodic and one
which satisfies the lesser condition of being "irreducible". In the
latter, the system will eventually sample all states starting from a given
initial state\cite{sokal}, but,
{\em at a fixed later time}\/, the probability for some of
the states is zero. A familiar example which is irreducible but not ergodic is
the Ising model at infinite temperature simulated
using Metropolis updating, for which the
probability to flip is unity in this limit. Clearly after an odd time, the
number of flipped spins must be odd and vice-versa. For such a non-ergodic 
system, it is possible for $P_m/\peq_m$ and $P_n/\peq_n$ to be different 
for states which have no common descendant at any fixed later time (for the 
Ising model example given here, states which differ by an odd number of spin
flips).

Note that Eq.~(\ref{dg_res}) does not give an
estimate for {\em how fast}\/ equilibrium is reached. 

For random updating considered so far
the probability of making a transition is the same for
every move, \ie\ writing Eq.~(\ref{master}) as
\begin{equation}
P_l(t+1) = \sum_m \Gamma_{lm} P_m(t),
\end{equation}
then $\Gamma$, the transition matrix (related to $w$ by
$\Gamma_{lm} = w_{m \rightarrow l}$),
is the same for each ``time'' $t$.
%Note that for $l\ne m$, $\Gamma_{lm} = w_{m \rightarrow l}$.
However, for sequential updating, the transition matrix depends on which site is
being updated, so, for a complete sweep, we have
\begin{equation}
\Gamma = \Gamma^{(1)}  \Gamma^{(2)} \cdots \Gamma^{(N)} ,
\label{composite}
\end{equation}
where $\Gamma^{ (i)}$ is the transition matrix for updating spin $i$.
Although the $\Gamma^{ (i)}$ individually satisfy the
detailed balance condition, the transition matrix for the whole sweep,
$\Gamma$, does {\em not},\cite{sokal}
%though it does preserve $\peq$ as a stationary distribution,
because the probability
of the reverse transition, $m\to l$ say, for a whole sweep,
is related to the probability of
transition $l\to m$ in the desired way only if the spins
are updated in the reverse order.
Despite overall lack of detailed balance, convergence to the equilibrium
distribution is still obtained for sequential updating because 
$G$ decreases at {\em each} step, as
long as each of the
transition probabilities, $\Gamma^{ (i)}$, satisfies the
detailed balance condition.

More generally, detailed balance is not necessary, see
Ref.\onlinecite{sokal} and references therein. In the rest of this paper, we
give a derivation of the necessary and sufficient conditions
for convergence to $\peq$: 
\begin{enumerate}
\item[ (i)]
the algorithm has $\peq$ as a stationary distribution
\item[ (ii)]
for any pair of states $(i,j)$, there exists some $T_{ij}$ and some 
state $k_{ij}$ such that at the $T_{ij}$'th time step the probabilities to have 
reached $k$ from $i$ and $j$ are both non-zero.
\end{enumerate}
We shall see that (ii) (with (i)) implies ergodicity. 
%The proof given here is somewhat similar to that in Ref.\onlinecite{K1}, but proceeds 
%forward in time; Ref.\onlinecite{K1} uses the transpose of the transition matrix,
%which does not correspond to any physical process, but loosely speaking moves
%backwards in time.

We first prove that conditions (i) and (ii) are sufficient.
Note that Eq.~(\ref{master})
is linear in the probabilities $P_l,$ so that if we define $\delta P_l = P_l
-\peq_l,$ then (with condition (i)) $\delta P_l$ also satisfies Eq.~(\ref{master}),
with $\sum_l \delta P_l = 0.$ 

It is convenient to use a measure of the deviation from equilibrium that is 
different from Eq.~(\ref{quaddev}):
\begin{equation}
L = \sum_l |\delta P_l|.
\label{lindev}
\end{equation}
Like $G,$ the quantity $L$ is positive unless the distribution has 
converged to $\peq.$ We denote by $W_{l\rightarrow m}(t)$ the probability 
that the system, starting out in state l, reaches state m after t time steps.
This is obtained by `iterating' the transition rates $\{w\}.$ Then 
\begin{equation}
L(t) = \sum_m |\sum_l W_{l\rightarrow m}(t)  \delta P_l(0)|\leq \sum_{m, l}
W_{l\rightarrow m}(t)|\delta P_l(0)|.
\label{evolution}
\end{equation}
Using the result $\sum_m W_{l\rightarrow m}(t) = 1, $ we see that $L(t)\leq 
L(0).$  We now compare the two sides of the inequality in Eq.(\ref{evolution}),
by choosing a specific value of $m$ and carrying out the $l$-summation. If
all states $l$ for which $W_{l\rightarrow m}\neq 0$ have the same sign for 
$\delta P_l(0)$, it is clear that the $l$-summation on both sides are equal.
On the other hand, if some of these states have $\delta P(0)>0$ and others have 
$\delta P(0) <0,$ the $l$-summation on the left hand side has both positive
and negative terms, and must be less than the corresponding sum on the right
hand side. Thus so long as there is at least one state $m$ which receives
`contributions' from two states $(i, j)$ with opposite $\delta P(0)$, 
i.e. $\delta P_i(0)$ and $\delta P_j(0)$ have opposite signs and 
$W_{i\rightarrow m}\neq 0$ and $W_{j\rightarrow m}\neq 0,$ we see
that $L(t) < L(0).$ 
Condition (ii) ensures
that for {\it any} $(i, j)$ this is the case for $t = T_{ij}.$ Thus $L(t)$
stays constant until $t = \min[T_{ij}],$ where the minimum is taken over all 
$(ij)$ for which $\delta P_i$ and $\delta P_j$ have the opposite sign, and then
decreases at that time step. The time $t$ can be reinitialized to zero at this 
point, and the whole argument repeated again. Note that nothing in the argument
requires the transition rates $w_{l\rightarrow m}$ to be time independent, so
long as conditions (i) and (ii) are always satisfied. Also, as with the first
approach above, no estimate has been
obtained for {\it how fast\/} $L(t)$ approaches zero. 

It may seem surprising that $L(t)$ need not decrease at every time step,
whereas Eq.~(\ref{dg_res}) shows that $G$ must. If we start out with $\delta P$
positive and negative on two states that are well separated from each other
(in the sense that many time steps are required before the two states have
common descendants), it is clear that $L(t)$ does not change at first.
However, $G$ decreases because the positive and negative $\delta P$'s both get
`smeared out' over several states.  Thus different measures can depict the
approach to equilibrium at different rates. Since any physical observable $O$
is given by $\langle O\rangle =\sum_l O_l P_l,$ the error $\langle\delta
O\rangle \leq \sum_l |O_l\delta P_l| \leq L \max_l[|O_l|], $ so that $L$ is a
conservative indicator of how observables approach equilibrium.

We have proved the sufficiency of conditions (i) and (ii); the necessity can
be easily demonstrated. The necessity of (i) is obvious. If (ii) is violated,
starting with an initial condition $\delta P_i(0) = -\delta P_j(0)$ and all 
other $\delta P_l$'s equal to zero, the positive and negative regions for 
$\delta P$ stay separate for all time and cannot neutralize each other.

Condition (ii) (with (i)) is equivalent to the 
(seemingly stronger) condition of ergodicity.
%\begin{enumerate}
%\item[(iii)] For sufficiently large $t,$ $W_{i\rightarrow j}(t)\neq 0$ for all
%pairs of states $(i, j)$
%\end{enumerate}
%(unless the state $j$ has $\peq_j = 0$).
If the system were not ergodic,  then starting with $P_l(0)=\delta_{il}$ would lead 
to a $P(t)$ with $P_j(t) = 0$ for some $j,$ since all states would not be accessible from
the state $i$ at time $t.$ Therefore $P(t)$ could not have converged to $\peq.$
Since we have seen that (i) and (ii) imply convergence, they must imply ergodicity.

The proof of convergence in Ref.\onlinecite{K1} is comparable in simplicity to
the proof given above. However, it applies the {\it transpose\/} $\Gamma^T$ of the 
transition matrix to the initial state of the system (represented as a column 
vector), and shows that under iteration the initial state evolves to a column 
vector whose entries are all identical. Although it is possible~\cite{K1} to 
deduce the desired properties of $\Gamma$ from this, there is no physical 
interpretation of evolution under $\Gamma^T,$ which does not even conserve
probability. In contrast, the proof given above shows how fluctuations 
from the desired distribution are "mixed'' by time evolution, with positive 
fluctuations annihilating negative ones. It is also physically clear why
ergodicity is essential for this annihilation process to proceed to completion:
positive and negative fluctuations are then assured of encountering each other
under time evolution.

Although we have considered the conditions for convergence to $\peq,$ in 
practice Monte Carlo simulations are not carried out using such an ensemble 
average, but by simulating a single system and taking a time average of 
many measurements. If we restrict ourselves to the case when the transition 
matrix $\Gamma$ is time-independent, this defines a homogeneous Markov 
chain~\cite{foot1}. It can then be shown that if the Markov chain is 
irreducible and has $\peq$ as a stationary distribution, the time average 
$\langle P(t)\rangle_t$ converges to $\peq.$ We indicate how to prove this 
result here: since $L(t)\leq L(0)$ in general,
none of the eigenvalues of $\Gamma$ can have modulus 
greater than unity~\cite{foot3}.
Irreducibility ensures\cite{foot2} that the eigenvalue 1
has a unique eigenvector, $\peq$.
The time averaging removes contributions from
oscillatory eigenvalues of $\Gamma,$ of the form $e^{i\theta}$ ($\theta\neq
0$).  By comparison, the stronger condition of ergodicity (ii) (with (i)) is
enough to rule out oscillatory eigenvalues, since no time averaging is needed
for convergence to $\peq.$ It is not clear how
this  proof that irreducibility is sufficient would generalize to time
dependent transition rates, since the evolution of $P$ is then not directly related
to the eigenvalues of $\Gamma(t)$.

We thank M.~E.~J.~Newman for useful correspondence.


\begin{references}
\bibitem{foot1}
    This includes the possibility of cycling periodically through a set of
    transition matrices, since the composite matrix defined in
    Eq.(\ref{composite}) is time independent, although it need not satisfy
    detailed balance even if its constituents do.
\bibitem{foot3}
    If the vectors of $\Gamma$ do not span the vector space, let $\{V_i\}$ be
    the eigenvectors with eigenvalues $\{v_i\}.$ Then it is possible to
    construct a $C_1$ in the complement of the eigenspace, such that $\Gamma
    C_1 = \lambda C_1 + \sum_i d_i V_i.$ By taking a suitable linear
    combination of $C_1$ with the $V_i$'s, one can find a $C_1^\prime$ such
    that $\Gamma C^\prime_1 = \lambda C_1 + \sum_i d_i^\prime V_i,$ where
    $d_i^\prime\neq 0$ {\em only\/} if $\lambda = v_i.$ Therefore $\Gamma^N
    C^\prime_1 = \lambda^N C^\prime_1 + N \lambda^{N-1} \sum_i d_i^\prime
    V_i.$  The non-increasing property of $L(t)$ then implies $|\lambda| < 1.$
    The process of constructing $C_i$'s can be iterated to show that the
    entire complement of the eigenspace of $\Gamma$ is annihilated by
    $\lim_{N\rightarrow\infty}\Gamma^N.$
\bibitem{foot2}
    Otherwise, a suitable linear combination of the several eigenvectors would
    yield an eigenvector $P^\prime$ whose elements would not all have the same
    sign; since $\Gamma^N P^\prime = P^\prime $ for all $N,$ by the discussion
    after Eq.(\ref{evolution}) this would require $\Gamma^N$ not to connect
    the positive and negative elements of $P^\prime$ for any $N,$ which would
    contradict the assumption that $\Gamma$ is irreducible.
\end{references}
\end{document}